\documentclass[aps,preprint,a4paper]{revtex4-1}
\usepackage{graphicx}
\usepackage{amsmath}
\usepackage{amssymb}
\usepackage{times}
\usepackage{ulem}
\usepackage[utf8]{inputenc}
\usepackage{placeins}
\usepackage{makecell}
\usepackage[version-1-compatibility]{siunitx}
\usepackage[pdfborder=0]{hyperref}
\hypersetup{
    colorlinks=true,
    linkcolor=blue,          
    citecolor=blue           
}

\def\nn{\nonumber}
\def\beq{\begin{equation}}
\def\eeq{\end{equation}}
\def\beqna{\begin{eqnarray}}
\def\eeqna{\end{eqnarray}}
\def\bea{\begin{array}}
\def\ea{\end{array}}

\def\mf{{\mathcal F}}

\def\MU{{\mathcal U}}
\def\MV{{\mathcal V}}
\def\mw{{\mathcal W}}

\begin{document}
\title{Stochastic resonance and amplification in the ac
driven Duffing oscillator with added noise}
\author{Adriano A. Batista}
\email{adriano@df.ufcg.edu.br}
\affiliation{
Departamento de Física\\
Universidade Federal de Campina Grande\\
Campina Grande-PB,
CEP: 58109-970, Brazil}
\author{A.~A. Lisboa de Souza}
\affiliation{
Departamento de Engenharia Elétrica, Universidade Federal da Paraíba\\
João Pessoa-PB, CEP: 58.051-970, Brazil}
\author{Raoni S. N. Moreira}
\affiliation{
Departamento de Física\\
Universidade Federal de Pernambuco\\
Recife-PE\\
CEP: 50670-901, Brazil}
\date{\today}
\begin{abstract}
Stochastic resonance (SR) is a coherence enhancement effect due to noise
that occurs in periodically-driven nonlinear dynamical systems.
A very broad range of physical and biological systems present this effect such
as climate change, neurons, neural networks, lasers, SQUIDS, and tunnel diodes,
among many others.
Early theoretical models of SR dealt only with overdamped bistable oscillators.
Here, we propose a simple model that accounts for SR in an underdamped 
driven Duffing oscillator with added white noise.
Furthermore, we develop a theoretical method to predict the effect of white noise
on the pump, signal, and idler responses of a Duffing amplifier.
We also calculate the power spectral density of the response of the 
Duffing amplifier.
This approach may prove to be useful for assessing the robustness
of acoustic, phononic, or mechanical frequency-comb generation to the
presence of noise.
\end{abstract}
\maketitle
\section{Introduction}
Stochastic resonance (SR) is a coherence enhancement effect caused by noise
that occurs in periodically-driven nonlinear dynamical systems.
The hallmark feature of this effect is characterized by the resonant behavior of
the signal-to-noise ratio as a function of the noise level.
A very broad range of physical and biological systems present this effect such
as climate change \cite{benzi1981mechanism}, neurons
\cite{longtin1993stochastic}, neural networks \cite{ikemoto2018noise}, 
lasers \cite{mcnamara1988observation}, 
SQUIDS \cite{hibbs1995stochastic}, tunnel diodes
\cite{mantegna1994stochastic}, 
nanomechanical oscillators \cite{badzey2005coherent}, etc.

Just a few years after the discovery of SR, in the mid 80's,
Jeffries and Wiesenfeld started studying the effect of small signals on
non-autonomous nonlinear dynamical systems near bifurcation points
\cite{jeffries1985observation, wiesenfeld1985noisy}.
They showed that several different systems are very sensitive to
noise and coherent perturbations near the onset of codimention-one bifurcations,
such as period doubling, saddle node, transcritical, Hopf, and pitchfork
(symmetry-breaking) bifurcations.
Initially, the effect of broadband noise was investigated (theoretically and
experimentally) near
period-doubling and Hopf bifurcations in a periodically driven $p$-$n$ junction
.
They found precursors of the bifurcations, such as new lines in the power
spectrum, as the noise level was increased.
Later on, a general theoretical framework, based on perturbation and Floquet
theories, explaining the effects of small coherent signals perturbing nonlinear
systems near the onset of bifurcation points was developed \cite{wies85,wies86},
with applications to the ac-driven Duffing oscillator.
It was found that nonlinear dynamical systems could be used as
narrow-band phase sensitive amplifiers.

With the advent and development of MEMS technology in the 90's, new
mechanical resonators were developed, such as the doubly-clamped beam
resonators that could reach very high quality factors.
The dynamics of the fundamental mode of these resonators is
well approximated by the Duffing equation.
Furthermore, these micromechanical devices exhibit a bistable response
that can be quantitatively modelled by Duffing oscillators
\cite{aldridge05}.
Nanomechanical resonators were implemented experimentally that use bifurcation
points and SR as a very sensitive means of amplification
\cite{almog07}.
Another method of amplification of small signals with a driven Duffing
oscillator with added noise was recently proposed by Ochs et al.
\cite{ochs2021amplification}.

Early theoretical models of SR \cite{mcnamara1989theory,
dykman1995stochastic, gammaitoni1998stochastic} dealt only with overdamped
bistable oscillators, neglecting altogether inertia terms in the dynamics.
Although other methods have been proposed to explain SR behavior in underdamped
oscillators, such as the method of moments \cite{alfonsi2000intrawell,
kang2003observing, landa2008theory}, in which a Fokker-Planck equation
model was used, the theory is considerably more complex than the model we
propose here.
Stocks {\it et al.} \cite{stocks1993stochastic} proposed a model for SR in an
underdamped monostable Duffing oscillator based on susceptibility calculations
obtained directly from the fluctuation-dissipation theorem
\cite{landau1980statistical} and the noise spectral density given in
\cite{dykman1990noise}.
When these pieces of the theory are combined the model becomes complex as
well.
Furthermore, one does not know a priori if it will work near bifurcation
points of the bistability region of the Duffing oscillator.
In addition, they did not consider the effects of noise on the Duffing
amplifier.
Here, we propose a simple model that accounts for SR in an underdamped 
forced Duffing oscillator with added white noise.
The oscillator could be dynamically bistable or monostable.
In what follows, we will focus on monostable SR.
Furthermore, we develop equations to predict the effect of white noise
on the pump, signal, and idler responses in a Duffing amplifier.
This approach may prove to be useful for assessing the effect of added noise
on the generation of frequency-comb spectra in mechanical resonators \cite{czaplewski2018bifurcation,
ganesan2017phononic, ganesan2018phononic, singh2020giant, batista2020frequency}.
We also calculate the power spectral density of the response of the driven
Duffing oscillator.
Here we extend previous works \cite{almog06, batista08} on driven Duffing
oscillators to include an analysis of amplification near bifurcation points
(such as near the cusp of the bistability region where two saddle-node
bifurcations merge) in the presence of noise.
Furthermore, we analyze the effects of noise on the bifurcation points and on
amplification, and we calculate signal-to-noise ratios (SNRs) with respect to
pump, idler, and signal responses.

The contents of this paper are organized as follows. 
In Sec. \ref{theory}, we review the ac-driven Duffing oscillator with damping
and the Duffing amplifier dynamical properties based on averaging techniques, on
the harmonic balance method, and on numerical integration of the equations of
motion.
In Sec. \ref{noise}, we develop a theoretical model to investigate stochastic
resonance due to added white noise on the driven underdamped Duffing oscillator.
In Sec. \ref{amplifierNoise}, we extend our model to investigate how the
amplification of pump, signal, and idler responses are affected by the added
white noise in the Duffing amplifier.
In Sec. \ref{conclusions}, we draw our conclusions.

\section{The Duffing oscillator and the Duffing amplifier}
\label{theory}
The dynamics of the Duffing amplifier is described by the following equation
\begin{equation}
    \ddot x+\omega_p^2x=-\Omega x-\gamma\dot x-\alpha x^3+F_p\cos(\omega_p
    t)+F_s\cos(\omega_s t+\phi_0),
    \label{amp}
\end{equation}
where $\Omega=1-\omega_p^2$ and
$\omega_s$ is the external signal frequency.
We suppose $\alpha=O(\epsilon)$, with $0<\epsilon<<1$.
We also assume that
$\Omega, \gamma,\, F_p>>F_s$ are all $O(\epsilon)$.
We then rewrite this equation in the form
$\dot x = y$, $\dot y = -\omega_p^2x+g(x,y,t)$,
where $ g(x,y,t)= -\Omega x-\gamma y-\alpha
x^3+F_p\cos(\omega_pt)+F_s\cos(\omega_p t+\phi(t))$.
Here, $\phi (t)=\delta t+\phi_0$, in which $\delta=\omega_s-\omega_p$.
We now set the above equation in slowly-varying form with the transformation to a slowly-varying frame
\beq
\left(\bea{c}
x \\ y
\ea
\right)
=
\left(
\begin{array}{cc}
    \cos\omega_p t & -\sin \omega_p t\\
    -\omega_p\sin\omega_p t & -\omega_p\cos\omega_p t
\end{array}
\right)
\left(\bea{c}
\MU \\ \MV
\ea
\right)
\label{slow_trans}
\eeq
   and obtain 
\beq
\left(\bea{c}
\dot \MU \\ \dot \MV
\ea
\right)
=
-\frac{1}{\omega_p}\left(\bea{c} \sin(\omega_pt) g(x,y,t)\\
\cos(\omega_pt) g(x,y,t)\ea\right).
\label{slow_frame}
\eeq

\begin{figure}[h!]
    \centerline{\includegraphics[{scale=0.5}]{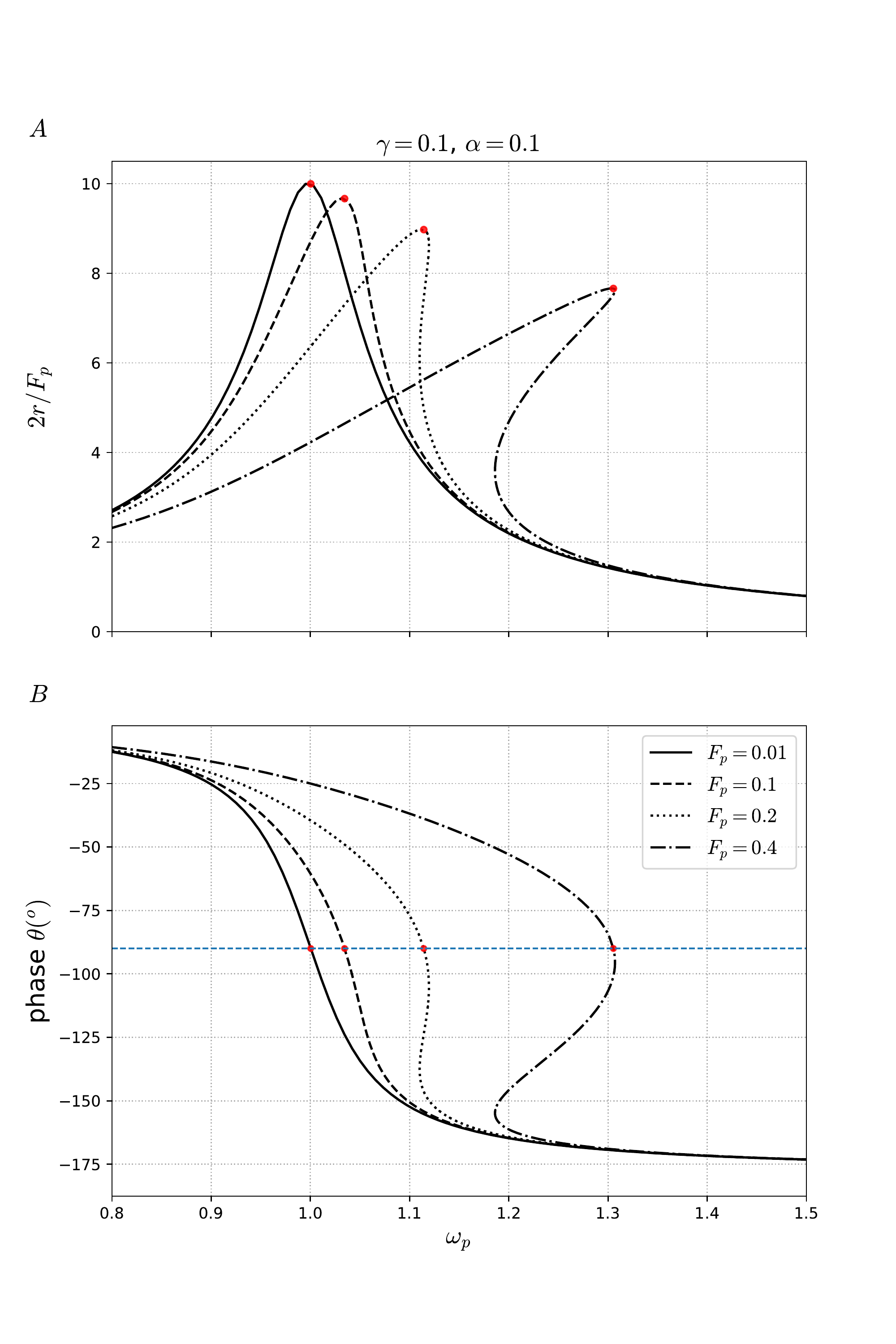}}
\caption{ $\mathbf A$ Nonlinear resonant curves as obtained from
Eq.~\eqref{roots} via numerical continuation (see Appendix A). 
The scaled response amplitude ($2r/F_p$) is represented as a function of the
angular pump frequency $\omega_p$. 
For low pump amplitude, the peak of the resonant curve tends to the quality
factor $Q=\gamma^{-1}=10$, consistent with the linear oscillator resonant curve.
As the pump amplitude increases, the nonlinear effects become more relevant.
The red dots are analytical approximations of the resonant peaks.
$\mathbf B$ Duffing oscillator response phase dependence on the drive frequency
$\omega_p$ as obtained from Eq.~\eqref{phase}.
The horizontal dashed line corresponds to the phase angle delay $-90^\circ$
of the oscillator response with respect to the driving force.}
\label{fig:BistableResonantCurve}
\end{figure}

After application of the AM to first order (in which, basically,
we filter out oscillating terms at $2\omega_p$ and $4\omega_p$ in the above
equation), we obtain
\beq
\begin{aligned}
    \dot{u} &= \frac{-1}{2\omega_p}\left[\gamma\omega_p  u
    +\Omega  v+3\alpha (u^2+v^2)v/4-F_s\sin\phi(t)\right],\\
    \dot { v} &= \frac{-1}{2\omega_p}\left[-\Omega  u+\gamma\omega_p v-3\alpha (u^2+v^2)u/4+F_p+F_s\cos\phi(t)\right],
\end{aligned}
    \label{1stOrdAvg}
\eeq
where the functions $\MU(t)$ and $\MV(t)$ are replaced by their slowly-varying
averages $u(t)$ and $v(t)$.
We are interested in the stationary solution of Eq.~\eqref{1stOrdAvg}.
If there is no external signal, $F_s=0$, the fixed points of the above equation can be found by solving the
following cubic in $r^2$
\beq
r^2\left[\left(\Omega+3\alpha r^2\right)^2+\gamma^2\omega_p^2\right]=F_p^2/4,
\label{roots}
\eeq
where $\bar u=2r\cos\theta$ and $\bar v=2r\sin\theta$.
From the above equations we find that the necessary conditions for the existence
of three real roots is $\alpha>0$ and $\Omega^2-3\gamma^2\omega_p^2>0$, or
$|\Omega|>\sqrt{3}\gamma\omega_p$ (i.e. 
$\left|\omega_p^2-1\right|> \sqrt{3}\gamma\omega_p$).
At the cusp point of the bistability region,
one obtains $\Omega^2=3\gamma^2\omega_p^2$ and $r^2=-2\Omega/(9\alpha)$, hence
the pump frequency has to be blueshifted in relation to the natural frequency of
the oscillator, that is $\omega_p>1$.
Furthermore, $F_p^2=32\sqrt{3}\gamma^3\omega_p^{*3}/(27\alpha)$,
where $\omega_p^*=\frac{\sqrt{3}\gamma+\sqrt{3\gamma^2+4}}{2}$.
The phase angle $\theta$ between pump drive and oscillator response can be found from 
\beq
\sin\theta=-\frac{2\gamma\omega_p r}{F_p}.
\label{phase}
\eeq
Due to the cubic nonlinearity in Eq.~\eqref{amp} with the coefficient $\alpha >0$, the elastic constant
increases with amplitude, hence, the elastic constant is larger than the
corresponding linear oscillator elastic constant.
As a consequence of this, one gets a shift of the resonant peak to a higher
frequency, a lower peak, and bistability as can be seen in
Fig.~\ref{fig:BistableResonantCurve}{\bf A}.
Note also that we can estimate the frequency and amplitude of the resonant peak.
From Eq.~\eqref{roots}, we obtain that the peak amplitude is
$2r_M\approx F_p/(\gamma\omega_M)$, where the resonant angular frequency is shifted to
approximately
\beq
\omega_M=\frac1{\sqrt{2}}\left(1+\sqrt{1+\frac{3\alpha F_p^2}{\gamma^2}}\right)^{1/2}.
\label{resonantFreq}
\eeq
We plot these points in both frames as red dots and we see that they pinpoint 
quite closely the resonant peaks.
In frame {\bf B}, we plot the phase $\theta$. 
One can see that at resonance the phase delay of the oscillator response
with respect to the driving force is $-90^\circ$.
This result remains the same in both the linear and the nonlinear regimes.

\FloatBarrier

\subsection{Analysis of the Duffing amplifier response to the signal}
We now perform a linear response analysis of the oscillator to the 
presence of the external signal, when $F_s\neq0$, in Eq.~(\ref{1stOrdAvg}). 
We use the following notation $u(t)=\bar u+u_1(t)$ and $v(t)=\bar v+v_1(t)$, where 
$\bar u$ and $\bar v$ are fixed-point solutions of Eq.~\eqref{1stOrdAvg} with
$F_s=0$. 
This results in
\begin{eqnarray}
    \dot { u_1} &=& \frac{-1}{2\omega_p}\left\{\gamma\omega_p  u_1
    +\Omega  v_1+3\alpha \left[2\bar u\bar vu_1+(\bar u ^2+3\bar v^2)v_1\right]/4-F_s\sin\phi(t)\right\},\nn\\
    \dot { v_1} &=& \frac{-1}{2\omega_p}\left\{-\Omega  u_1+\gamma\omega_p v_1-3\alpha \left[(\bar v^2+3\bar u^2)u_1+2\bar u\bar vv_1\right]/4+F_s\cos\phi(t)\right\}.
    \label{linearResponse}
\end{eqnarray}
Using the harmonic balance method, in which we assume
$u_1(t)=\tilde u_1e^{i\delta t}+\tilde u_1^*e^{-i\delta t}$
and $v_1(t)=\tilde v_1e^{i\delta t}+\tilde v_1^*e^{-i\delta t}$, we obtain
\begin{eqnarray}
     i\delta \tilde u_1 &=& \frac{-1}{2\omega_p}\left\{\gamma\omega_p  \tilde u_1
     +\Omega  \tilde v_1+3\alpha \left[2\bar u\bar v\tilde u_1+(\bar u ^2+3\bar v^2)\tilde v_1\right]/4+i F_se^{i\phi_0}/2\right\},\nn\\
     i\delta \tilde v_1 &=& \frac{-1}{2\omega_p}\left\{-\Omega  \tilde
     u_1+\gamma\omega_p \tilde v_1-3\alpha \left[(\bar
     v^2 +3\bar u^2) \tilde u_1+2\bar u\bar v\tilde v_1\right]/4+F_se^{i\phi_0}/2\right\}.
    \label{harmBalance}
\end{eqnarray}
This algebraic linear system may be recast as

\beq
\left(
\bea{c}
\tilde u_1\\
\tilde v_1
\ea
\right)
=
-\frac{F_se^{i\phi_0}}{4\omega_p(ad-bc)}
\left(
\bea{cc}
d & -b \\
-c & a
\ea
\right)
\left(
\bea{c}
i \\
1
\ea
\right),
\label{u1v1}
\eeq
where the coefficients are
\beqna
a&=&i\delta +\gamma/2+\frac{3\alpha\bar u\bar v}{4\omega_p},\nn\\
b&=&\frac{\Omega+3\alpha(\bar u^2+3\bar v^2)/4}{2\omega_p},\nn\\
c&=&-\frac{\Omega+3\alpha(\bar v^2+3\bar u^2)/4}{2\omega_p}\nn\\
d&=&i\delta +\gamma/2-\frac{3\alpha\bar u\bar v}{4\omega_p}.\nn
\eeqna
The response of the oscillator may be written approximately as
\beq
x(t)= \bar u\cos(\omega_p t)-\bar v\sin(\omega_p t)
+ u_s\cos[(\omega_p+\delta)t]-v_s\sin\left[(\omega_p +\delta)t\right]
+ u_i\cos[(\omega_p-\delta)t]-v_i\sin\left[(\omega_p -\delta)t\right],
\label{signalIdler}
\eeq
where $u_s=\mbox{Re} [\tilde u_1]-\mbox{Im} [\tilde v_1]$,
$v_s=\mbox{Im} [\tilde u_1]+\mbox{Re} [\tilde v_1]$,
$u_i=\mbox{Re} [\tilde u_1] +\mbox{Im} [\tilde v_1]$, and 
$v_i=-\mbox{Im} [\tilde u_1] +\mbox{Re} [\tilde v_1]$.
The terms at frequency $\omega_p$ correspond to the pump response,
the terms at $\omega_p+\delta$ are known as the signal response, and
the terms at $\omega_p-\delta$ are known as the idler response.

We can define the gains in decibels of the Duffing amplifier response with
respect to the signal excitation in decibels as
\beq
\begin{aligned}
G_p &= 20\log\frac{X_p}{F_s},\\
G_s &= 20\log\frac{X_s}{F_s},\\
G_i &= 20\log\frac{X_i}{F_s},
\end{aligned}
\label{gain_signal}
\eeq
where $X_p=\sqrt{u^2+v^2}$, $X_s=\sqrt{u_s^2+v_s^2}$, and
$X_i=\sqrt{u_i^2+v_i^2}$.

\begin{figure}[h!]
    \centerline{\includegraphics[{scale=0.6}]{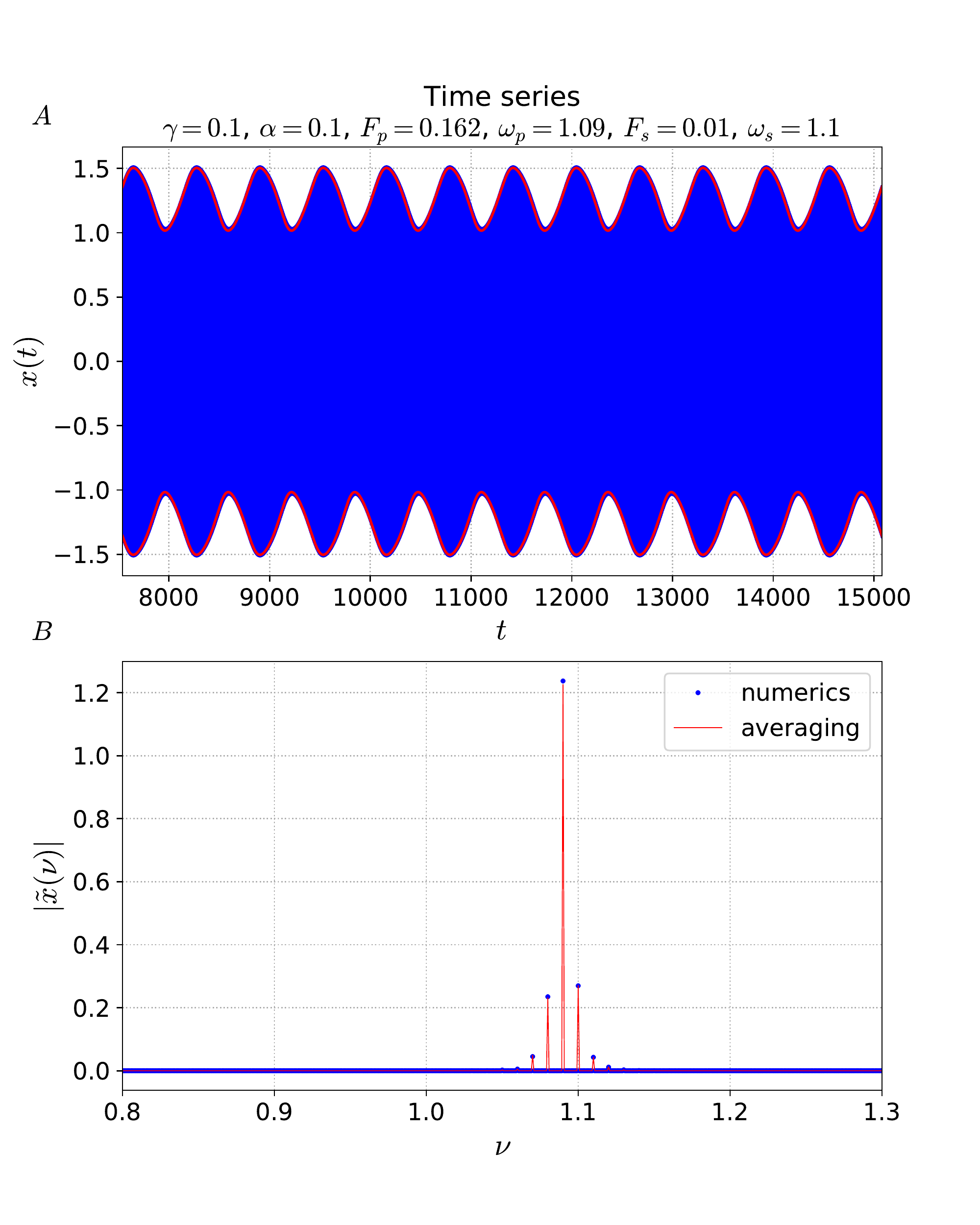}}
\caption{ {\bf A} Time series at the cusp point of the bistability region. 
The envelope is given by $\sqrt{u(t)^2+v(t)^2}$, which can be obtained from the
linear response result of Eq.~\eqref{1stOrdAvg}.
{\bf B} The corresponding Fourier transform. 
We again obtain good agreement between numerical results with averaging method predictions.
}
\label{fig:timeseries}
\end{figure}

\begin{figure}[h!]
    \centerline{\includegraphics[{scale=0.6}]{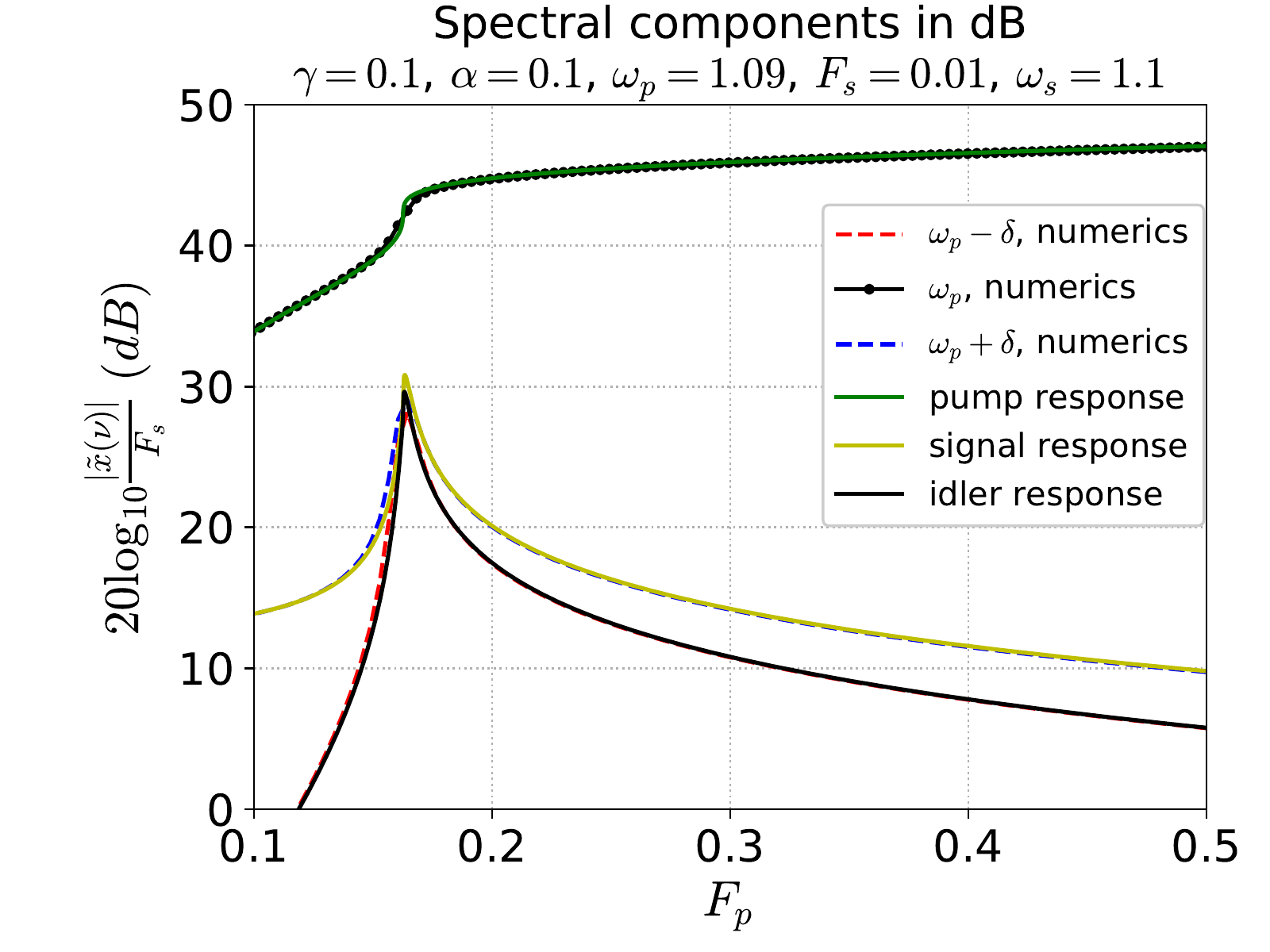}}
\caption{Spectral components as function of $F_p$.
Numerical and analytical (linear response) results are plotted together.
The analytical results are obtained from Eqs.~\eqref{u1v1}-\eqref{gain_signal}.
The peaks in the signal and idler gains occur near the saddle-node bifurcation 
which is at the cusp of the bistability region.}
\label{fig:peaks}
\end{figure}

In Fig.~\ref{fig:timeseries}, we show a time series of the numerical integration
of the Duffing amplifier equations of motion given in Eq.~\eqref{amp}. 
The envelope is obtained from the averaging method via Eq.~\eqref{1stOrdAvg}.
In Fig.~\ref{fig:peaks}, we compare the gains of pump, signal, and idler 
responses as defined in Eq.~\eqref{gain_signal}.
Very good agreement between numerical results and the linear response
predictions, as given in Eqs.~\eqref{u1v1}-\eqref{gain_signal}, are obtained.

\FloatBarrier

\FloatBarrier

\section{The underdamped ac-driven Duffing oscillator with added white noise}
\label{noise}
\subsection{Stochastic resonance}
\begin{figure}[h!]
    \centerline{\includegraphics[{scale=0.6}]{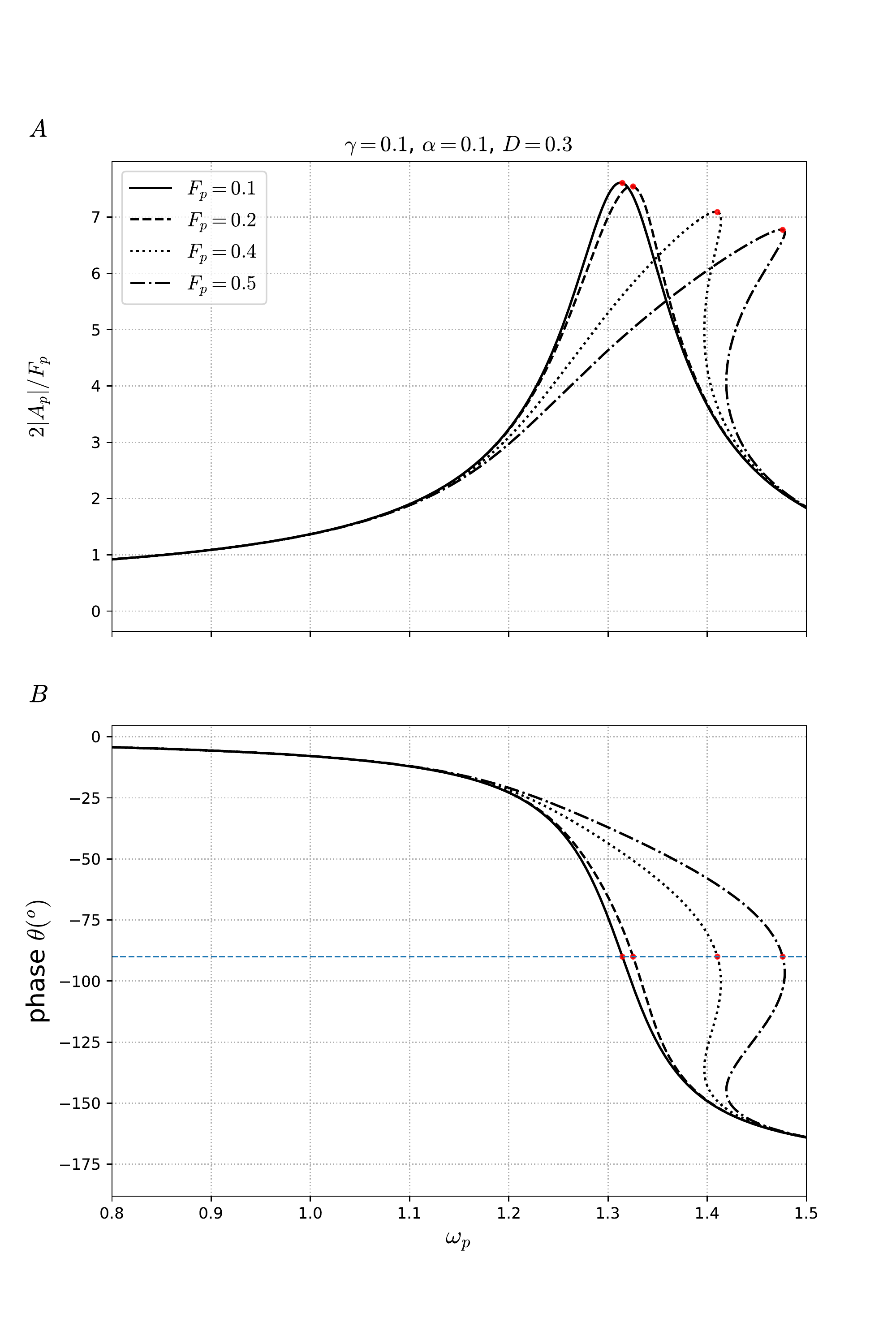}}
\caption{$\mathbf A$ Effect of noise on resonance curves.
The curves are blue shifted and the peaks are reduced when noise is added
with respect to the equivalent results, with $D=0$, in Fig.
\ref{fig:BistableResonantCurve}.
Furthermore, some curves with bistability, when $D=0$, are reduced or devoid of bistability when $D=0.3$. 
The red dots are analytical approximations of the resonant peaks.
$\mathbf B$ Duffing oscillator response phase dependence on the drive frequency
$\omega_p$ as obtained from Eq.~\eqref{eq:Aw}.
The horizontal dashed line corresponds to the phase angle delay $-90^\circ$
of the oscillator response with respect to the driving force.}
\label{fig:noiseBistab}
\end{figure}

\begin{figure}[h!]
    \centerline{\includegraphics[{scale=0.7}]{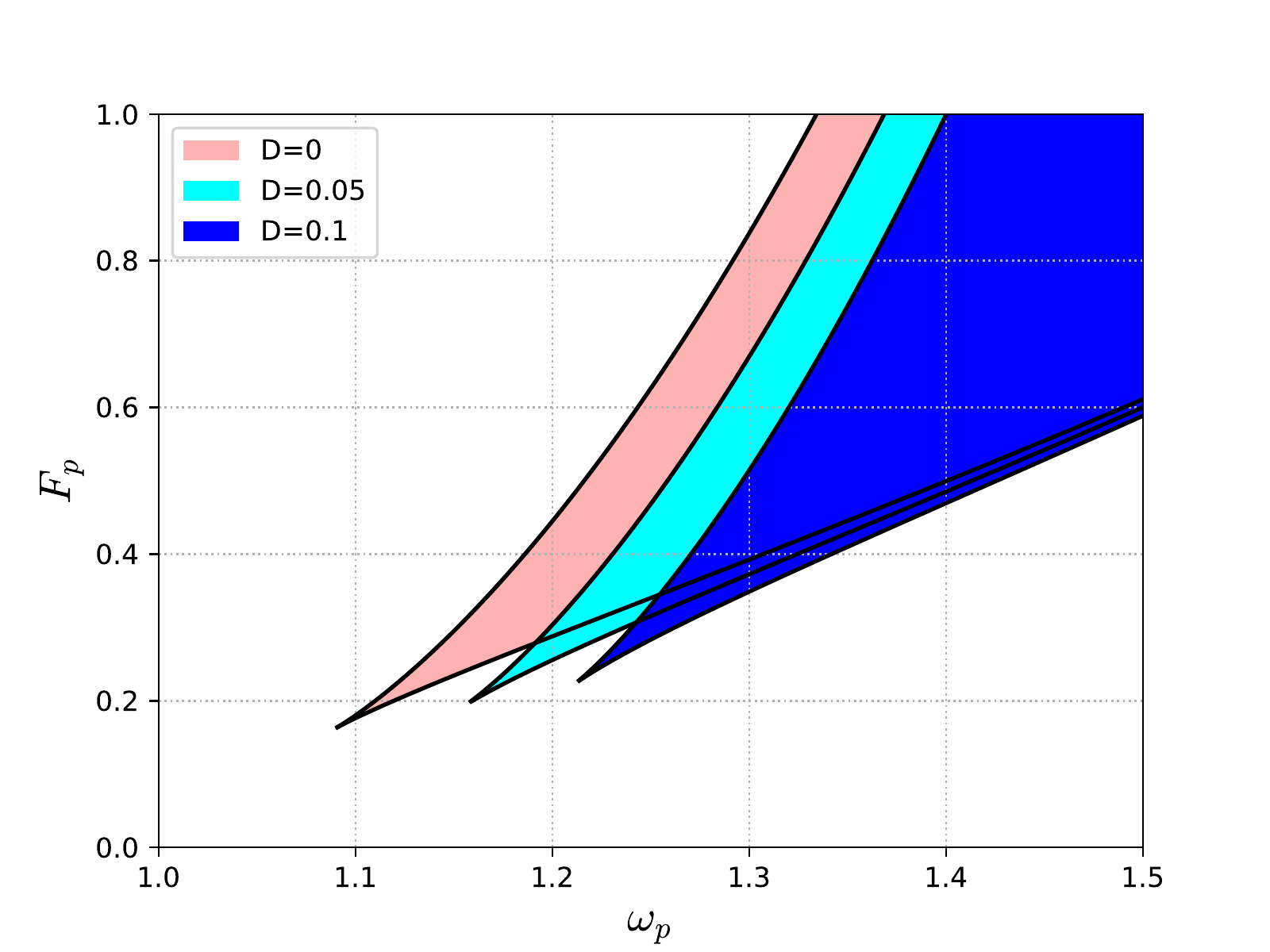}}
\caption{Comparison of bistability regions (shaded) of the Duffing oscillator 
without white noise, based on Eq.~\eqref{roots}, and with white noise, based on 
Eqs.~\eqref{stocRes}. 
The parameters are $\gamma=0.1$ and $\alpha=0.1$.}
\label{fig:bisRegionNoise}
\end{figure}

\begin{figure}[h!]
\centerline{\includegraphics[{scale=0.7}]{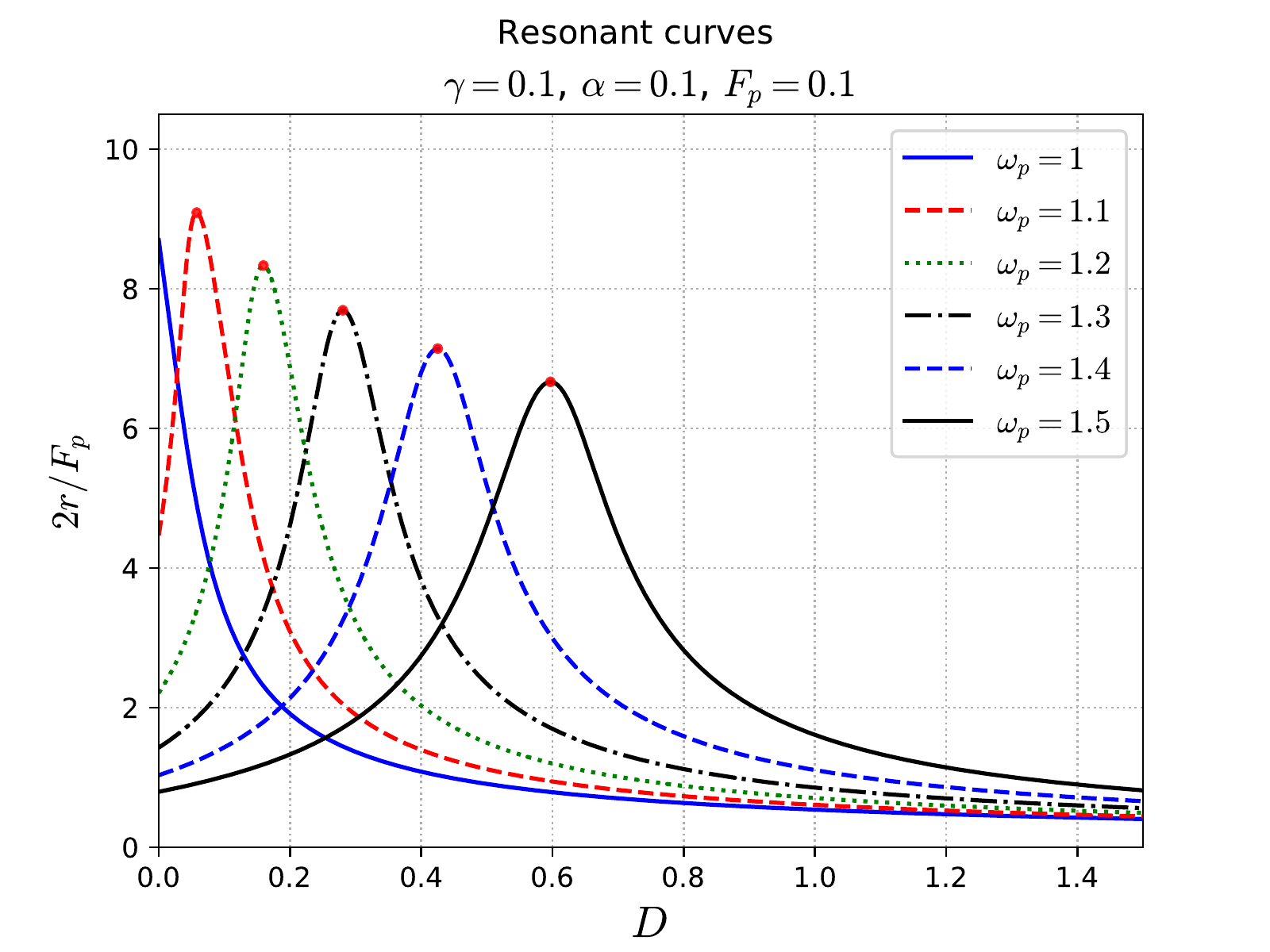}}
\caption{Resonance curves as a function of $D$. 
The curves are obtained from Eq.~\eqref{stocRes} of the
Duffing oscillator response to a pump with angular frequency $\omega_p$. 
The red dots are analytical approximations of the resonance peaks.
}
\label{fig:stocRes}
\end{figure}

The time evolution of the forced Duffing oscillator in the presence of
dissipation and noise is given by
\begin{equation}
    \ddot x+x=-\gamma\dot x-\alpha x^3+F_p\cos(\omega_p t)+R(t),
\end{equation}
where the Gaussian noise $R(t)$ obeys $\langle R (t)\rangle=0$ and $\langle R
(t)R(t')\rangle=2D\delta(t-t')$.
We then assume the response of the Duffing oscillator can be split in
two parts: one coherent and the other stochastic, such as
\beq
x(t)=A_p e^{i\omega_p t}+A^*_p e^{-i\omega_p t}+\delta x(t),
\eeq
where $\delta x(t)$ is the response to the input noise.
We obtain approximately the following differential algebraic system
\beqna
\delta \ddot x &=&  -\delta x -\gamma \delta \dot x-\alpha \left(6|A_p|^2+\overline{\delta x^2}_\infty\right)\delta x+R(t)\label{LangOsc},\\
A_p&=& \frac{F_p}{2\left[ \Omega+i\gamma \omega_p+3\alpha \left(|A_p|^2+\overline{\delta x^2}_\infty\right)
\right]},
\label{eq:Aw}
\eeqna
where we neglected the superharmonics and made the approximation 
$\delta x^3\approx \overline{\delta x^2}_\infty\delta x$ for the random 
fluctuations in Eq.~\eqref{LangOsc}.
The average $\overline{\delta x^2}_\infty$ is the equilibrium quadratic
fluctuation of 
the stochastic process deviate $\delta x$ and is given by
\[
\overline{\delta x^2}_\infty=\frac{D/\gamma}{1+\alpha\left(6 |A_p|^2+\overline{\delta x^2}_\infty\right)}.
\]
The stationary solution is given by
\beq
\begin{aligned}
\overline{\delta x^2}_\infty &=
\frac{-(1+6\alpha|A_p|^2)+\sqrt{(1+6\alpha|A_p|^2)^2+4\alpha D/\gamma} }{2\alpha},\\
|A_p| &= \frac{F_p}{2\sqrt{\left[\Omega+3\alpha
\left(|A_p|^2+\overline{\delta x^2_\infty}\right)\right]^2+\gamma^2\omega_p^2}}.
\end{aligned}
\label{stocRes}
\eeq
One can observe in Eq. \eqref{stocRes} that a back-action effect occurs
between the fluctuations due to noise and the oscillator fundamental harmonic
amplitude.
As a consequence, there is a shift of the resonance peak of the
oscillator pump response, i.e. $|A_p|$, to higher frequencies
as can be seen in Fig.~\ref{fig:noiseBistab} when compared to resonant curves
of Fig.~\ref{fig:BistableResonantCurve}{\bf A}. 
This shift increases with increasing noise level.
The red dots correspond closely to the resonant peaks in which 
$\Omega+3\alpha(|A_p|^2+\overline{\delta x^2_\infty})=0$.
In Fig.~\ref{fig:bisRegionNoise}, we see that the region of bistability
decreases and is displaced to higher frequencies and higher pump amplitudes.
This effect implies that stochastic resonance can only occur in our system when
the driving pump frequency is blue-shifted in relation to the natural frequency
of the oscillator, i.e.  $\Omega<0$, as can be seen in Fig.~\ref{fig:stocRes}.
Simply speaking, SR occurs when the bistability region passes by or near a
parameter-space point $(\omega_p, F_p)$ when the noise level $D$ is increased.
This back-action effect is due to the cubic nonlinearity of the Duffing
oscillator and to the fact that the squared deviate $\delta x^2(t)$ has a finite
average. 
We also would like to point out that as the noise level is increased the
bistability width in pump frequency decreases in qualitative agreement with
the experimental work by Aldridge and Cleland \cite{aldridge05}.
This is expected since bistability is a coherent response of the nonlinear
oscillator to the pump which decreases in amplitude and range when the noise
level increases. 

\FloatBarrier

\FloatBarrier
\subsection{Noise spectral density and signal-to-noise ratio}
\begin{figure}[h!]
\centerline{\includegraphics[{scale=0.6}]{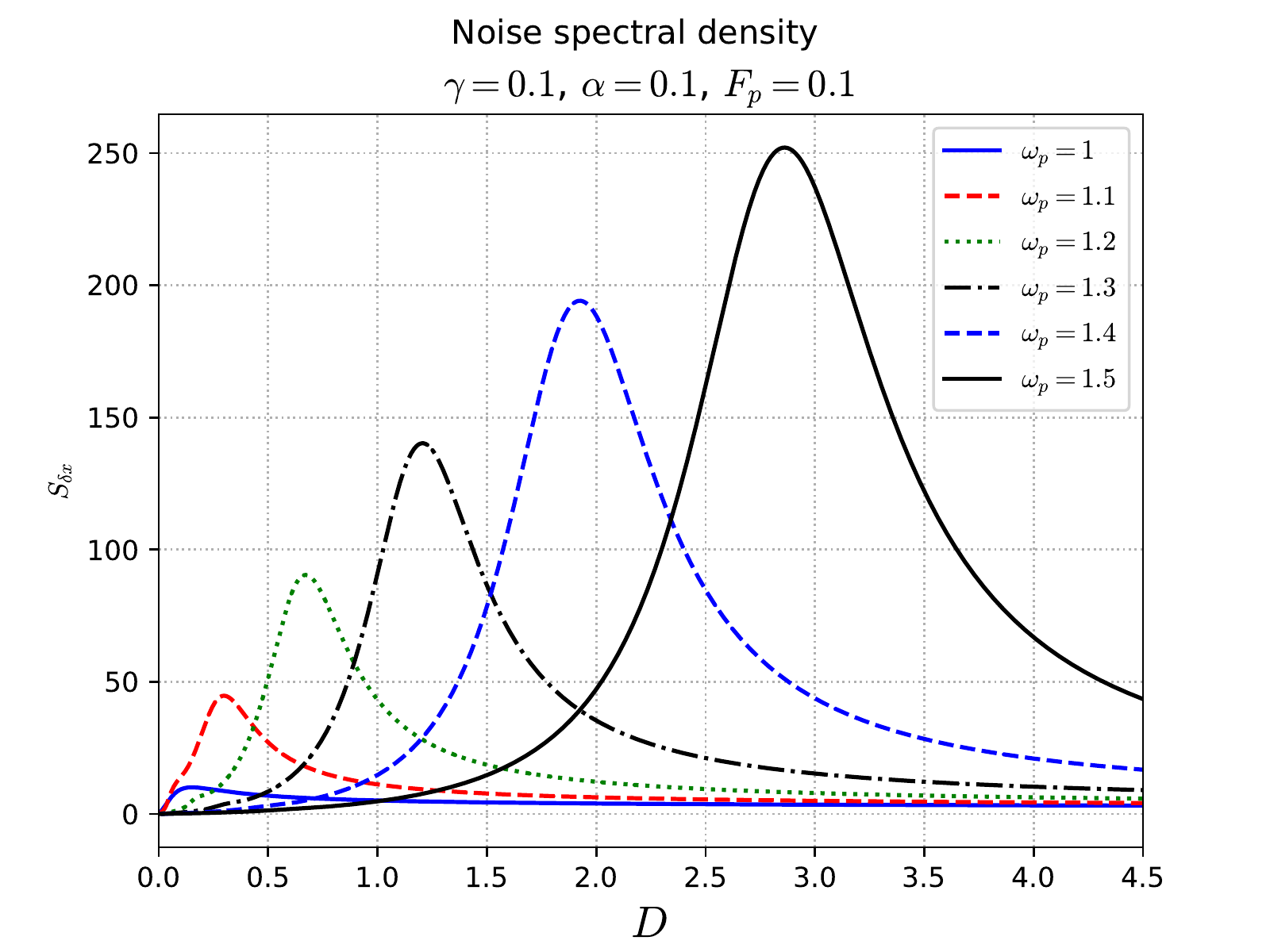}}
\caption{Noise spectral density curves 
obtained from Eq.~\eqref{eq:S_x} for
various values of pump frequency as a function of noise level $D$.
}
\label{fig:S_x}
\end{figure}

\begin{figure}[h!]
\centerline{\includegraphics[{scale=0.6}]{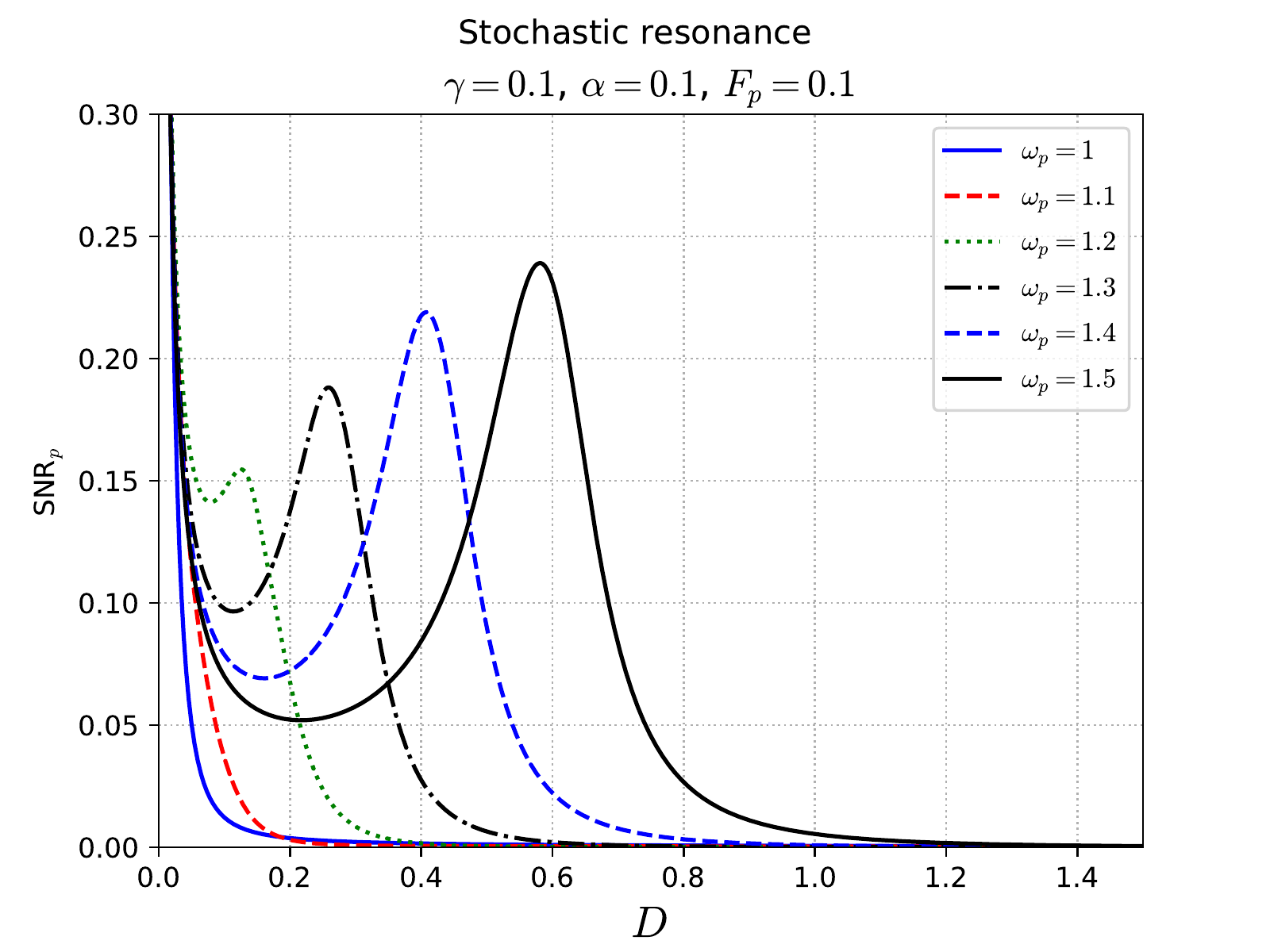}}
\caption{SNR curves obtained from Eq.~\eqref{eq:SNR} for various values of pump frequency as a function of noise level
$D$.
The peaks in these curves are characteristic of stochastic resonance.
}
\label{fig:SNR}
\end{figure}

With the appropriate modifications, the power spectral density of the stationary
stochastic process  $\delta x(t)$, as governed by
Eq.~\eqref{LangOsc}, can be written as
\beq
S_{\delta x}(\nu) =\int_{-\infty}^{\infty}\langle \delta x(t+\tau) \delta x(t)
\rangle\, e^{-i\nu\tau}d\tau =\frac{2D}{\left(\nu^2-\nu_0^2\right)^2+\gamma^2\nu^2},
\label{eq:S_x}
\eeq
where $\nu_0^2= 1+\alpha \left(6|A_p|^2+\overline{\delta
x^2_\infty}\right)$, $A_p$ and $\overline{\delta x^2_\infty}$ are 
determined in Eq.~\eqref{stocRes}.
The signal-to-noise ratio (SNR) is given by
\beq
SNR_p = \frac{4|A_p|^2}{S_{\delta x}(\omega_p)}
\label{eq:SNR}
\eeq
In Fig.~\ref{fig:S_x}, we show the noise spectral density for various values
of pump frequencies. 
We note that the peaks here are detuned compared to the corresponding resonance
peaks of the pump response amplitude $|A_p|$ from
Fig.~\ref{fig:stocRes}.
There are no units for $S_{\delta x}$ or $D$ because we nondimensionalized
our equations.
In Fig.~\ref{fig:SNR}, we can see stochastic resonance in the SNR as a function
of the noise level $D$. 
We notice that it only occurs when the oscillator is driven by a blue-shifted
pump.
Furthermore, the effect grows with increasing detuning up to roughly
$\omega_p=2.35$ (not shown here).
After that the peaks in SNR start reducing.
In addition, for such high detuning, the noise level $D$ necessary for the
appearance of SR is considerably higher.

\FloatBarrier
\section{The Duffing amplifier with added white noise}
\label{amplifierNoise}
We now proceed to evaluate  estimates of signal-to-noise ratio (SNR) of the system described by the following dynamics
\begin{equation}
    \ddot x+x=-\gamma\dot x-\alpha x^3+F_p\cos(\omega_p
    t)+F_s\cos(\omega_s t+\phi)+R(t).
    \label{langevin2}
\end{equation}
We start our analysis of this stochastic differential equation seeking
a stationary solution of the type
\beq
x(t)= x_p(t)+x_s(t)+x_i(t)+\delta x(t),
\eeq
in which the coherent responses are
\[
\begin{aligned}
    x_p(t) &= A_p e^{i\omega_p t}+ A_p^* e^{-i\omega_p t},\\
    x_s(t) &= A_s e^{i\omega_s t}+A_s^* e^{-i\omega_s t},\\
    x_i(t) &= A_i e^{i\omega_i t}+ A_i^* e^{-i\omega_i t}.
\end{aligned}
\]
After removing the coherent evolution part from Eq.~\eqref{langevin2}, we 
obtain the following Langevin equation for the stochastic variable $\delta
x(t)$
\beq
\delta\ddot x   +\delta x +\gamma \delta \dot x+\alpha
\left[6\xi_0^2+\overline{\delta
x^2}_\infty\right]\delta x=R(t), 
\label{eq:noiseResponse2}
\eeq
where 
\beq
\xi_0^2=|A_p|^2+|A_s|^2+|A_i|^2.
\label{eq:xi02}
\eeq
The average of the stationary stochastic variable $\delta x(t)$ squared is
given by
\[
\overline{\delta x^2}_\infty =\frac{D/\gamma}{1+\alpha\left(6
 \xi_0^2+\overline{\delta x^2}_\infty\right)},\nn
\]
whose solution in terms of $\xi_0^2$ is
\begin{equation}
\overline{\delta x^2}_\infty =
\frac{-(1+6\alpha\xi_0^2)+\sqrt{(1+6\alpha\xi_0^2)^2+4\alpha
D/\gamma} }{2\alpha}.
\label{eq:dx2xi2}
\end{equation}
Using the harmonic balance method in Eq.~\eqref{langevin2}, we obtain
the algebraic system
\beq
\begin{aligned}
G_pA_p&=F_p/2-6\alpha A_iA_sA_p^*=
F_p/2+18\alpha^2|A_s|^2|A_p|^2A_p/G_i,\\
G_sA_s&=-3\alpha A_i^*A_p^2+F_se^{i\phi}/2=9\alpha^2|A_p|^4A_s/G_i^*+F_se^{i\phi}/2,
\\
G_iA_i&=-3\alpha A_s^*A_p^2. 
\end{aligned}
\label{PSIN_response}
\eeq
We also used the following shorthand notation 
\beq
\begin{aligned}
G_p(\omega_p, |A_p|^2, |A_s|^2, |A_i|^2) &= 
\Omega_p+3\alpha\left(|A_p|^2+2|A_s|^2+2|A_i|^2+\overline{\delta
x^2_\infty}\right)+i\gamma\omega_p,\\
G_s(\omega_s, |A_p|^2, |A_s|^2, |A_i|^2) &=
\Omega_s+3\alpha\left(2|A_p|^2+|A_s|^2+2|A_i|^2+\overline{\delta
x^2_\infty}\right)+i\gamma\omega_s,\\
G_i(\omega_i, |A_p|^2, |A_s|^2, |A_i|^2) &=
\Omega_i+3\alpha\left(2|A_p|^2+2|A_s|^2+|A_i|^2+\overline{\delta
x^2_\infty}\right)+i\gamma\omega_i,
\end{aligned}
\label{eq:GpGsGi}
\eeq
in which the detunings are given by $\Omega_p=1-\omega_p^2$, $\Omega_i=1-\omega_i^2$, and $\Omega_s=1-\omega_s^2$.

We can simplify the algebraic system given in Eqs.~\eqref{PSIN_response} and
obtain the equations for $|A_p|^2$, $|A_s|^2$, and $|A_i|^2$, which are 
given by
\beq
\begin{aligned}
|A_p|^2&=\frac{F_p^2}{4\left|G_p-18\alpha^2|A_s|^2|A_p|^2/G_i\right|^2},\\
|A_s|^2&=\frac{F_s^2}{4\left|G_s-9\alpha^2|A_p|^4/G_i^*\right|^2},\\
|A_i|^2&=9\alpha^2 |A_s|^2|A_p|^4/|G_i|^2. 
\end{aligned}
\label{eq:Awsi}
\eeq

\begin{figure}[h!]
    \centerline{\includegraphics[scale=1.0]{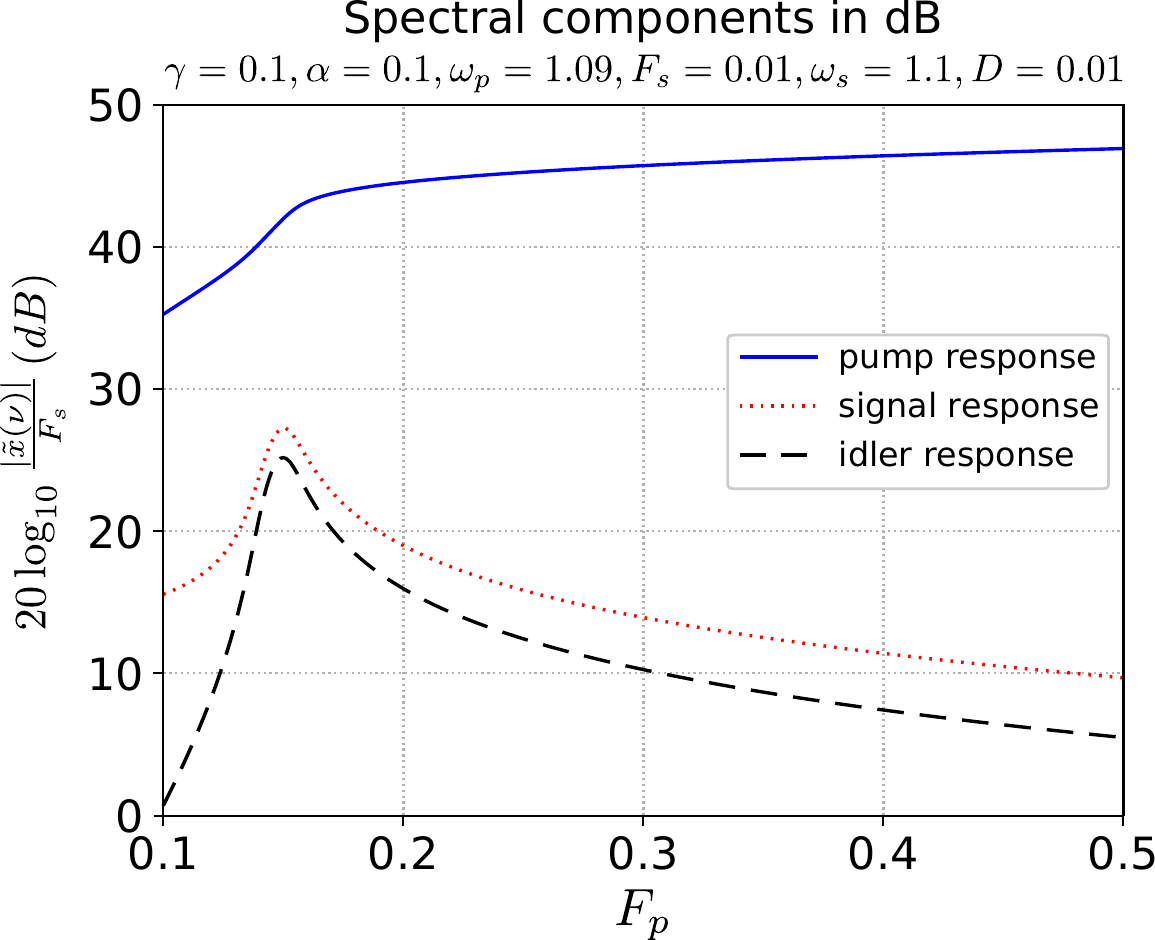}}
\caption{Main spectral components of the Duffing amplifier as a function of
$F_p$ in the presence of noise.
These  results are obtained from numerically solving Eqs.~\eqref{eq:Awsi}
for each value of the noise level $D$.
The peaks in the signal and idler gains occur near the cusp point
of the bistability region.
As the noise level increases, the bistability region becomes further away in
parameter space, as can be seen in Fig.~\ref{fig:bisRegionNoise}, hence the
amplification decreases as compared with the results presented in Fig.
\ref{fig:peaks}.
}
\label{fig:peaksD}
\end{figure}

In Fig.~\ref{fig:peaksD}, we plot the main spectral components of the Duffing
amplifier as a function of $F_p$ in the presence of noise.
These results are obtained from numerically solving the system of equations
\eqref{eq:dx2xi2} and \eqref{eq:Awsi} using the method 
of Gauss-Newton \cite{nocedal2006numerical}.
Notice that the peaks in the signal and idler gains are reduced compared
to the case in which there is no noise, $D=0$, as portrayed in
Fig.~\ref{fig:peaks}.
This occurs because due to the presence of noise the bistability region
moves further away in parameter space, as can be seen in
Fig.~\ref{fig:bisRegionNoise}, hence the amplification decreases.

\begin{figure}[h!]
\centerline{\includegraphics[{scale=0.6}]{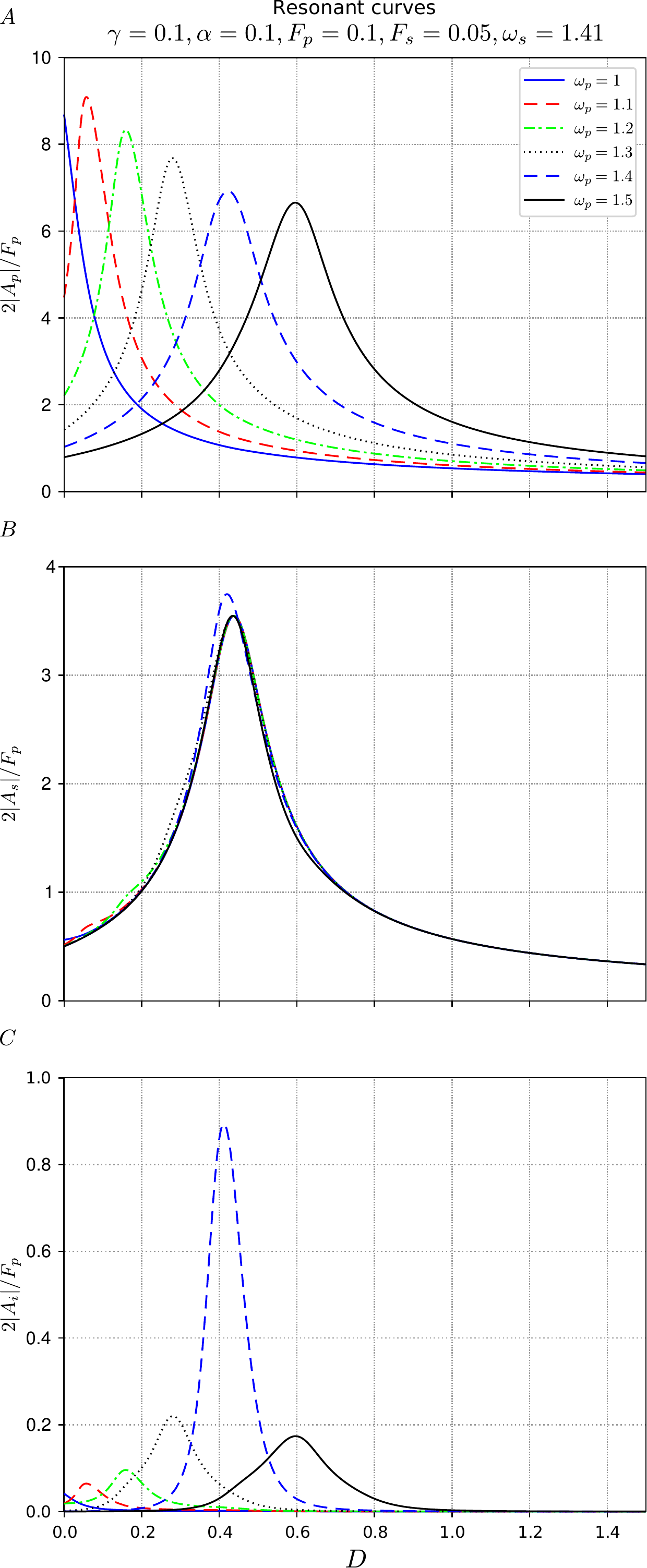}}
\caption{Duffing amplifier coherent response resonance curves as a function of 
the noise level $D$. 
The curves are obtained from Eq.~\eqref{eq:Awsi} for the Duffing amplifier
response to the simultaneous pump and signal coherent excitations and also
to the added white noise.
In frame $\mathbf A$, we show the pump responses to several values of pump
frequency.
In frame $\mathbf B$, we show the signal responses and in frame $\mathbf C$,
the idler responses.
In all responses we see clearly the occurrence of stochastic resonance.
}
\label{fig:Awsi}
\end{figure}

In Fig.~\ref{fig:Awsi}, we plot the main spectral components of the Duffing
amplifier scaled by $F_p$ as a function of the noise level $D$.
These results are obtained from numerically solving the system of equations
\eqref{eq:dx2xi2} and \eqref{eq:Awsi}.

\FloatBarrier
\subsection{Noise spectral density and signal-to-noise ratios}
\begin{figure}[h!]
\centerline{\includegraphics[{scale=0.6}]{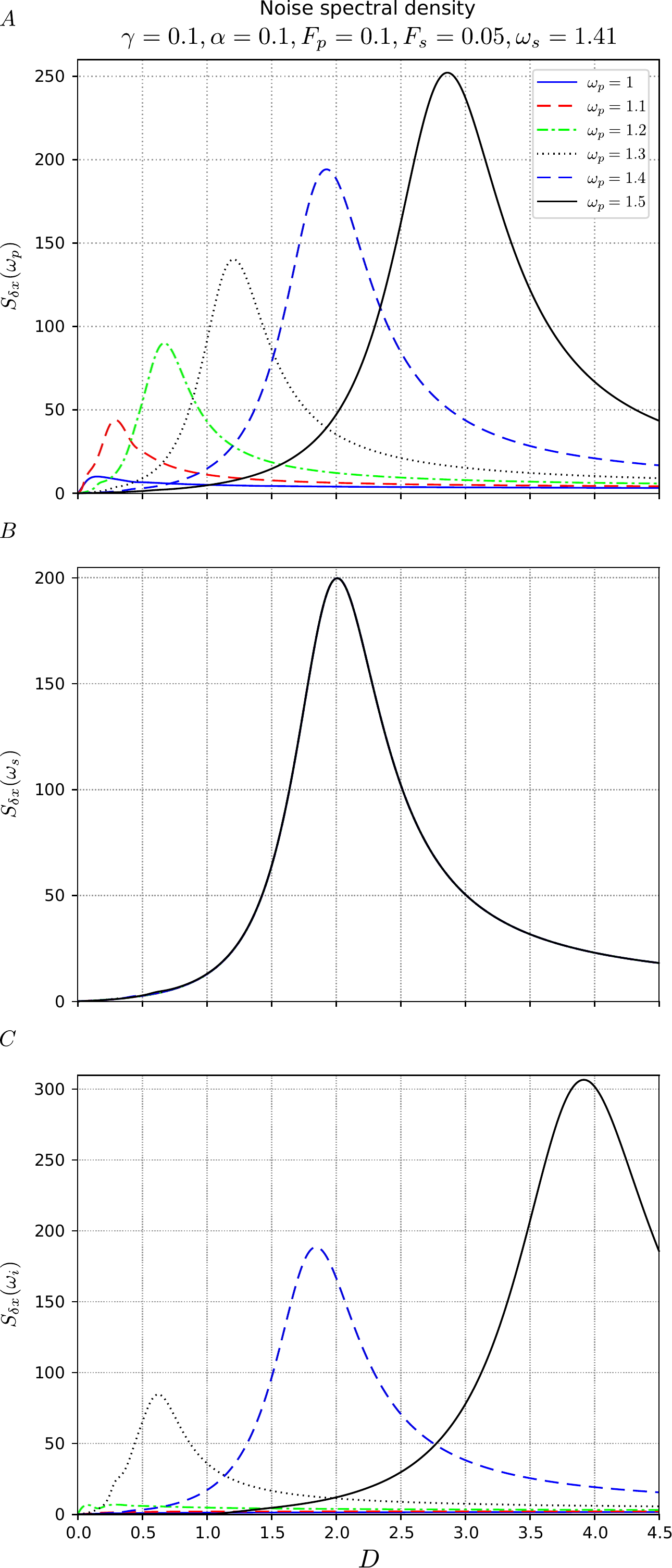}}
\caption{Noise spectral density curves at pump, signal, and idler frequencies 
obtained from Eq.~\eqref{eq:AmpS_x} for
various values of pump frequency as a function of noise level $D$.
}
\label{fig:AmpS_x}
\end{figure}

\begin{figure}[h!]
\centerline{\includegraphics[{scale=0.6}]{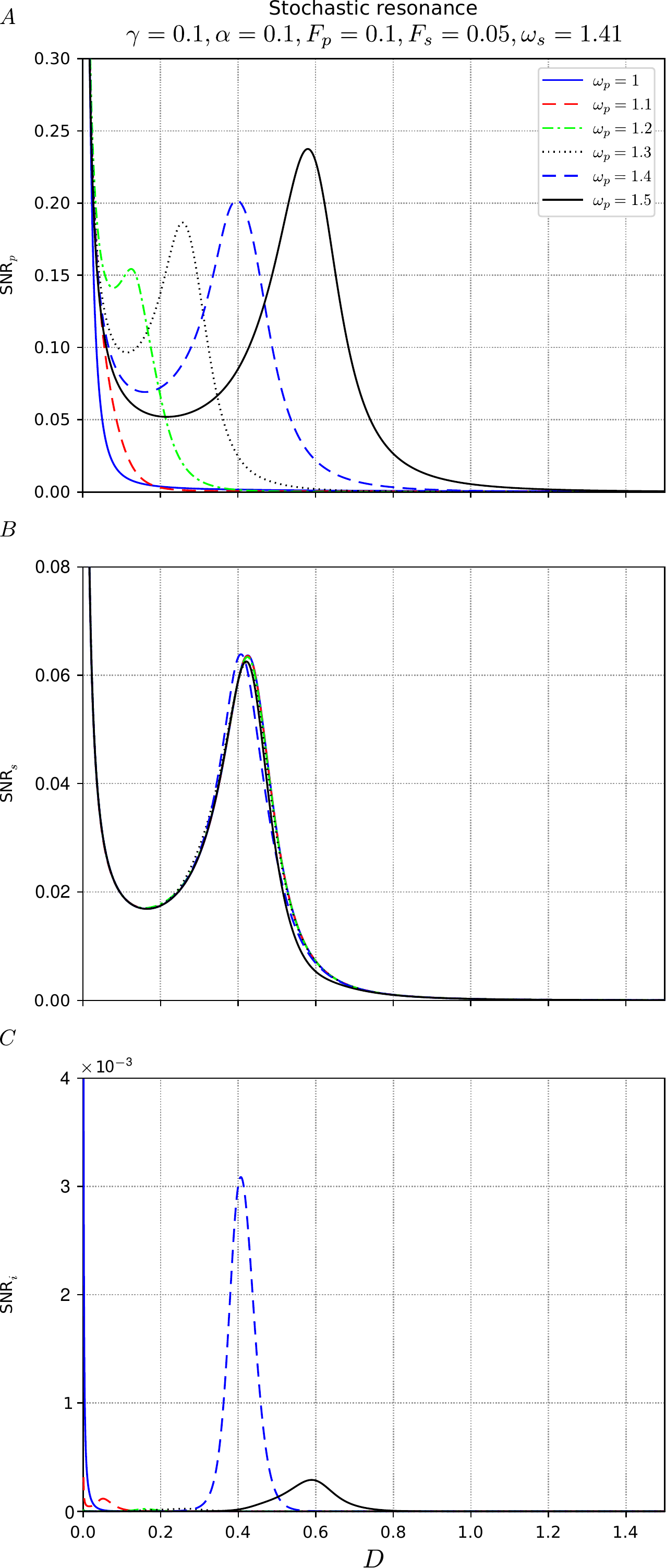}}
\caption{SNR curves obtained from Eq.~\eqref{eq:SNRs} for various values of pump frequency as a function of noise level
$D$.
}
\label{fig:SNRs}
\end{figure}

With the appropriate modifications, the power spectral density of the stationary
stochastic process  $\delta x(t)$, as governed by Eq.~\eqref{eq:noiseResponse2}, can be written as
\beq
S_{\delta x}(\nu) =\int_{-\infty}^{\infty}\langle \delta x(t+\tau) \delta x(t)
\rangle\, e^{-i\nu\tau}d\tau =\frac{2D}{\left[\nu^2-1-\alpha
\left(6\xi_0^2+\overline{\delta
x^2_\infty}\right)\right]^2+\gamma^2\nu^2},
\label{eq:AmpS_x}
\eeq
where $\xi^2$ is determined in Eq.~\eqref{eq:xi02} and 
$\overline{\delta x^2_\infty}$ by Eq.~\eqref{eq:dx2xi2}.
We can obtain then three measures of signal-to-ratio (SNR), 
each one related to pump, signal, and idler responses.
They are
\beq
\begin{aligned}
    SNR_p &= \frac{4|A_p|^2}{S_{\delta x}(\omega_p)},\\
    SNR_s &= \frac{4|A_s|^2}{S_{\delta x}(\omega_s)},\\
    SNR_i &= \frac{4|A_i|^2}{S_{\delta x}(\omega_i)}.
\end{aligned}
\label{eq:SNRs}
\eeq
In Fig.~\ref{fig:AmpS_x}, we show the noise spectral density (NSD) for various
values of pump frequencies for pump, signal, and idler frequencies. 
There is considerable dispersion in the peaks of pump and idler NSDs, but not
in the signal NSD.
We note that the peaks here are detuned compared to the corresponding resonance
peaks of the pump response amplitude $|A_p|$ from
Fig.~\ref{fig:Awsi}.
There are no units for $S_{\delta x}$ or $D$ because we nondimensionalized
our equations.
In Fig.~\ref{fig:SNRs}, we can see stochastic resonance in the SNR at pump
($\mathbf A$), signal ($\mathbf B$), and idler ($\mathbf C$) frequencies as a function of the noise
level $D$. 

\FloatBarrier

\section{Conclusion}
\label{conclusions}
Here, we briefly reviewed the theory on the dynamics of the ac-driven Duffing
oscillator with damping and the Duffing amplifier.
Afterwards, we investigated the effect of added white noise on the driven
Duffing oscillator and on the Duffing amplifier.
We proposed a simple approximate theoretical model to account for the effects
of coherent drive and added noise on the nonlinear system that goes beyond the linear response theory.
We assumed that the response of the nonlinear system to these inputs can be
split into coherent and stochastic parts that influence on one another.
We predict a blue shift in frequency (when the Duffing constant is positive)
and also a shift towards higher pump values of the bistability region that are
due to increased noise levels. 
For a constant value of pump amplitude, the range of bistability tends to
decrease with increasing noise levels. 
This was observed experimentally by Aldridge and Cleland \cite{aldridge05} and
is in qualitative agreement with our results.
A more quantitative agreement could be found if one uses the Euler-Bernoulli
beam theory and the parameters of their resonator to obtain the coefficients
of the Duffing oscillator. 
An outline of this method can be seen in the Appendix B.
In addition to these results, with our proposed model, we showed that the
coherent response and the noise spectral density have resonance peaks at
very different noise levels.
Due to this, there is a peak in SNR indicating that SR occurs.
We point out that the observed SR is related to the shifting of the
bistability region as the noise level is increased.
We have seen that in our system a necessary condition for SR to occur
is that the pump frequency has to be blue-shifted with respect to the
natural frequency of the nonlinear resonator. 
Furthermore, we notice that our model predicted SR in underdamped monostable
driven Duffing oscillators, but we emphasize that it could also be used to
investigate SR in under- or over-damped bistable Duffing oscillators.

We also predicted with semi-analytical methods that SR occurs 
in the pump, signal, and idler responses of the Duffing amplifier.
For each of these cases, we calculated the coherent response amplitude squared
and the NSD.
Again, as the coherent amplitude peaks occur at very different values of noise
levels than the NSD's peaks, a peak in the SNR appears and this is
characteristic of SR.
We also saw that the pump, signal, and idler SR peaks usually occur at different
values of noise levels.

We believe that the theoretical framework we developed here to obtain the
response to noise in the driven Duffing oscillator and in the Duffing amplifier
can be adapted to other nonlinear oscillators and amplifiers.
Quantitative predictions should be possible, once one obtains physical 
parameters from micro or nanomechanical resonators.
For example, one could investigate how the noise spectral
density that was measured in a parametrically-driven resonator by Miller {\it et
al.} \cite{miller2020spectral} will change when nonlinearities are present.
Another issue in  
parametric amplifiers is the presence of nonlinear dissipation
\cite{papariello2016ultrasensitive, li2020effects}.
How these systems behave when white noise is added to them seems to be an open problem. 

Particularly relevant systems to be investigated for the effect of added
noise are nonlinear resonators that can present frequency-comb spectra 
\cite{ganesan2017phononic, ganesan2018phononic, batista2020frequency}.
We believe that the method we developed here may help evaluate the robustness
of mechanical frequency-comb generation in the presence of noise.

\appendix
\section{Numerical continuation}
We solve numerically Eq.~\eqref{roots} for a range of pump frequency values
using the method of numerical continuation.
We call
\[
h(x, \omega_p)= x\left[\left(\Omega+3\alpha x\right)^2+\gamma^2\omega_p^2\right]-F_p^2/4,
\]
We notice that the dynamics
\beq
\begin{aligned}
    \dot x &= -\frac{\partial h}{\partial\omega_p},\\
    \dot\omega_p & = \frac{\partial h}{\partial x},
    \label{ham_flow}
\end{aligned}
\eeq
keeps $h(x(t), \omega_p(t))$ constant.
This is a Hamiltonian dynamics where $h$ is the Hamiltonian.
In the present case, the partial derivatives are given by
\[
\begin{aligned}
    \frac{\partial h}{\partial x} &= (\gamma\omega_p)^2+(\Omega+3\alpha
    x)^2+6\alpha x(\Omega+3\alpha x),\\
    \frac{\partial h}{\partial\omega_p} &= x\omega_p\left[-2(\Omega+3\alpha
    x)+\gamma^2\right].
\end{aligned}
\]
At $t=0$, we choose $h(x(0), \omega_p(0))=0$.
To avoid numerical difficulties, we normalize the flow in Eq.~\eqref{ham_flow}, such that the speed
$\sqrt{\dot x^2+\dot\omega_p^2}=1$.
\section{Single-degree-of-freedom theory}
Here, we show how to obtain a single-degree-of-freedom (SDOF) model from a
physical system such as a mechanical resonator, which could be a cantilever
or a doubly-clamped beam.
The simplest theory to describe the flexural vibrations of a thin beam is the
Euler–Bernoulli beam theory \cite{landau1986elasticity}. 
In this theory the bending motion is described by the equation
\beq
\mu\frac{\partial^2 X}{\partial t^2}=-EI\frac{\partial ^4X}{\partial z^4},
\label{eq:vibracoes}
\eeq
where $X(z, t)$ represents the lateral deflections from equilibrium at a point 
$z$ along the length of the beam at a time $t$.
In this equation, $\mu$ is the linear density of the beam (with units of mass
over length), $E$ is the Young's
modulus, and $I$ is the area moment of inertia and in a prismatic rod it
is given by 
\beq
I=\frac{\mw\tau^3}{12},
\label{momentoInercia}
\eeq
where $\mw$ is the width and $\tau$ is the thickness.
Using the method of separation of variables, we can write down the $n$-th
normal-mode solution of Eq.~\eqref{eq:vibracoes} as 
$X(z, t)=X_n(z)\cos(\omega_n t+\varphi)$, where $n=0, 1, 2,\dots$.
Hence, we obtain the differential equation for the amplitude $X_n(z)$, which is 
\beq
\frac{d^4X_n}{dz^4}=\frac{\mu\omega_n^2}{EI}X_n.
\label{eq:amplitude_z}
\eeq
The most general solution to this equation is
\beq
X_n(z)= A_n\cos(k_nz)+B_n\sin(k_nz)+C_n\cosh(k_nz)+D_n\sinh(k_nz),
\label{eq:X_0z}
\eeq
in which $k_n=\left(\frac{\mu\omega_n^2}{EI}\right)^{1/4}$.
The coefficients of this solution are determined by the boundary conditions.
Please see table \ref{resonators} for the main types of boundary conditions
used in nanomechanics.

\FloatBarrier
\begin{table}
\begin{center}
\begin{tabular}{cccc} 
Type of resonator &Boundary conditions & Characteristic equation &
Fundamental-mode eigenvalue\\
\hline
 \hline
 Clamped-clamped
 &\makecell{$X_n(0)=X_n^{'}(0)=0$\\$X_n(\ell_0)=X_n^{'}(\ell_0)=0$}&
$\cos(k_n\ell_0)\cosh(k_n\ell_0)=1$ &$4.730041$\\
\hline
 Clamped-free &\makecell{$X_n(0)=X_n^{'}(0)=0$\\
 $X_n^{''}(\ell_0)=X_n^{'''}(\ell_0)=0$}&
$\cos(k_n\ell_0)\cosh(k_n\ell_0)=-1$ &1.875104\\
\hline
\hline
\end{tabular}
\caption{Thin prismatic beam resonator boundary conditions with corresponding
characteristic equation and fundamental-mode eigenvalue of flexural vibrations.}
\label{resonators}
\end{center}
\end{table}
\FloatBarrier
From the $n$-th root $x_n$ of the characteristic equations in table
\ref{resonators}, 
we obtain
the normal mode frequencies of oscillation from the following expression
\begin{align}
f_n &= \frac{1}{2\pi}\sqrt{\frac{EI }{\mu}}k_n^2
= \frac{x_n^2\tau}{2\pi\ell_0^2}\sqrt{\frac{E}{12\rho}}.
\label{eq:f0_scaling}
\end{align}

The potential elastic energy  of the bent bar \cite{landau1986elasticity} can
be written as
\[
U_{el}=\frac{EI}2\int\kappa(s)^2 ds,
\]
where $s$ is the arclength parameterization and $\kappa(s)$ is the curvature.
In terms of $X(z)$, it can be written as
\[
\kappa(s)=\frac{X''(z)}{(1+X'(z)^2)^{3/2}},
\]
while $ds= \sqrt{1+X'(z)^2}dz$.
Hence, we can write the potential energy approximately as a
single-degree-of-freedom (SDOF) polynomial as
\begin{align}
&\frac{\kappa(z)X^2(z,t)}{2}+\frac{A(z)X^4(z,t)}{4}
\simeq\frac{EI}{2}\int_0^{\ell_0}\frac{X''^2}{\left(1+X'^2\right)^{5/2}}dz'\\\nn
&\approx\frac{EI}{2}\int_0^{\ell_0}X''^2\left[1-\frac{5}{2}X'^2\right]dz',
\end{align}
where the exact energy functional was approximated by a Taylor expansion of
the denominator.
We now use the first normal mode approximation $X(z, t) = X_0(z)\cos(\omega t)$
to find analytical expressions for the coefficients.
Based on the orthogonality condition of the Fourier expansion terms
$\cos(n\omega t)$, we are able to determine the coefficients above.
Hence, we find the coefficients of the SDOF potential energy polynomial to be
given by
\begin{subequations}
\begin{align}
    \kappa(z) &= \frac{EI}{X_0(z)^2}\int_0^{\ell_0}X_{0}''(z')^2dz'=
    \frac{EIx_0^4}{X_0(z)^2\ell_0^4}\int_0^{\ell_0}X_{0}(z')^2dz',
    \label{eq:kappa}\\
    A(z) &=
    -\frac{5EI}{X_0(z)^4}\int_0^{\ell_0}\left[X_0''(z')X_0'(z')\right]^2dz'
    \label{eq:alpha}
\end{align}
\end{subequations}
Using the first normal-mode solution, Eq.~\eqref{eq:X_0z}, we find that the
effective mass for the fundamental normal mode at height $z$
\cite{batista2018flexural} is
\beq
m_0^{eff}(z)=\frac{\mu}{X_0(z)^2}\int_0^{\ell_0}X_0(z')^2dz'.
\label{m_eff}
\eeq
The Newton's equation of motion for the SDOF approximation is given by
\begin{equation}
    m_{eff}\frac{d^2X}{ds^2}+\kappa X=-\Gamma\frac{dX}{ds}
    -A X^3+\mf_p\cos(\bar\omega_p s)+\mf_s\cos(\bar\omega_s s+\phi)+R(s).
    \label{langevin}
\end{equation}
In dimensionless units, we find Eq.~\eqref{langevin2}, where
the parameters are
\beq
\begin{aligned}
    \omega_0^2 &= \frac\kappa{m_{eff}},\\
    t &= \omega_0 s,\\
    x(t) &= \frac {X(t)}L,\\
    \omega_p &= \frac{\bar\omega_p}{\omega_0},\\
    \omega_s &= \frac{\bar\omega_s}{\omega_0},\\
    \gamma &= \frac{\Gamma}{m_{eff}\omega_0},\\
    \alpha &= \frac{AL^2}{m_{eff}\omega_0^2},\\
    F_p &= \frac{\mf_p}{m_{eff}\omega_0^2},\\
    F_s &= \frac{\mf_s}{m_{eff}\omega_0^2},\\
    r(t) &= \frac{R(s)}{m_{eff}\omega_0^2},
\end{aligned}
\label{eq:dless_pars}
\eeq
where $L$ is a characteristic length of the problem.
It should be chosen in such a way that the coefficient $\alpha$ is small,
otherwise the perturbative methods used in this paper may not work.
\FloatBarrier


%

\end{document}